\documentclass{ws-p9-75x6-50}

\begin{document}

\addtolength{\vsize}{1cm}

\title{Stable evolution of a maximally sliced Schwarzschild spacetime
using a smooth lattice.\footnote{A very brief report on the talk given in
parallel session CM1-a, {\em IX Marcel Grossman Meeting}, Rome, July 2000.}}

\author{Leo Brewin}

\address{
Department of Mathematics and Statistics, Monash University,
Clayton, 3800, Australia.
\\E-mail: Leo.Brewin@sci.monash.edu.au}

\maketitle

\abstracts{
We will present results of a long term stable evolution, to $t=1000m$, of a
maximally sliced Schwarzschild blackhole using a smooth lattice method.}

In recent papers~\cite{leo1,leo2} we have shown how local sets of Riemann normal
coordinates on a lattice can be used to generate numerical spacetimes. In this
talk we will report on a combination of the ADM equations with the smooth
lattice method to produce long term stable evolution of a maximally sliced
Schwarzschild blackhole.

A suitable lattice for a spherically symmetric space is a semi-infinite
ladder. The ladder has rungs of length $L_x$ and each rung is separated by
struts of length $L_z$. With zero shift, the ADM equations on the
lattice are
\def\lhs#1{\hbox to 0.9cm{\hfill$\displaystyle#1$}}
\def\tag#1{\hbox to 0.5cm{#1\hfill}}
\begin{eqnarray}
\tag{\hskip-1.0cm \hbox{\em Constraint equations}}&\nonumber\\
\lhs{0}&=&R_{xyxy} + 2R_{xzxz}+K_{xx}(K_{xx}+2K_{zz})\label{eq:ham}\\[3pt]
\lhs{0}&=&\frac{d}{dz}\left(L^2_xK_{zz}\right) - K_{xx}\frac{dL^2_x}{dz}\\[3pt]
\tag{\hskip-1.0cm \hbox{\em Evolution equations}}&\nonumber\\
\lhs{\frac{dL_x}{dt}} &=& -N K_{xx}L_x\\[3pt]
\lhs{\frac{dL_z}{dt}} &=& -N K_{zz}L_z\\[3pt]
\lhs{\frac{dK_{xx}}{dt}} &=& -\frac{1}{L_x} \frac{dL_x}{dz} \frac{dN}{dz} 
                                       + N (R_{xyxy}+R_{xzxz} + (2K_{xx}+K_{zz})K_{xx})\\[3pt]
\lhs{\frac{dK_{zz}}{dt}} &=& -\frac{d^2N}{dz^2} + N (2R_{xzxz} + (2K_{xx}+K_{zz})K_{zz})\\[3pt]
\tag{\hskip-1.0cm \hbox{\em Curvatures}}&\nonumber\\
\lhs{0}&=&\frac{d^2L_x}{dz^2} + L_x R_{xzxz}\label{eq:a}\\[3pt]
\lhs{0}&=&\frac{d}{dz}\left(L^2_xR_{xyxy}\right) - R_{xzxz}\frac{dL^2_x}{dz}\label{eq:b}\\[3pt]
\tag{\hskip-1.0cm \hbox{\em Maximal slicing}}&\nonumber\\
\lhs{0}&=&\frac{d^2N}{dz^2} + \frac{2}{L_x} \frac{dL_x}{dz} \frac{dN}{dz} - RN
\end{eqnarray}
The parameter $z$ is the cumulative proper distance measured along the radial
geodesic from the throat. Each of the $z-$derivatives were approximated by 2nd-order
accurate centred finite differences. In solving the constraints we
used the analytic value for $R_z$ with $m=1$ at the throat while the value for
$R_x$ at the throat was always set from the Hamiltonian constraint
(\ref{eq:ham}). Standard boundary conditions were applied, namely
reflection symmetry at the throat and static conditions at the outer boundary.
The evolution equations were integrated using a standard 4-th order Runge-Kutta
routine. The radial grid was chosen exactly as per
Bernstein, Hobill and Smarr~\cite{BHS}
with the one exception that we choose 800 rather than 400 grid points.

\begin{figure}[ht]
\begin{minipage}[t]{\textwidth}
    \epsfig{file=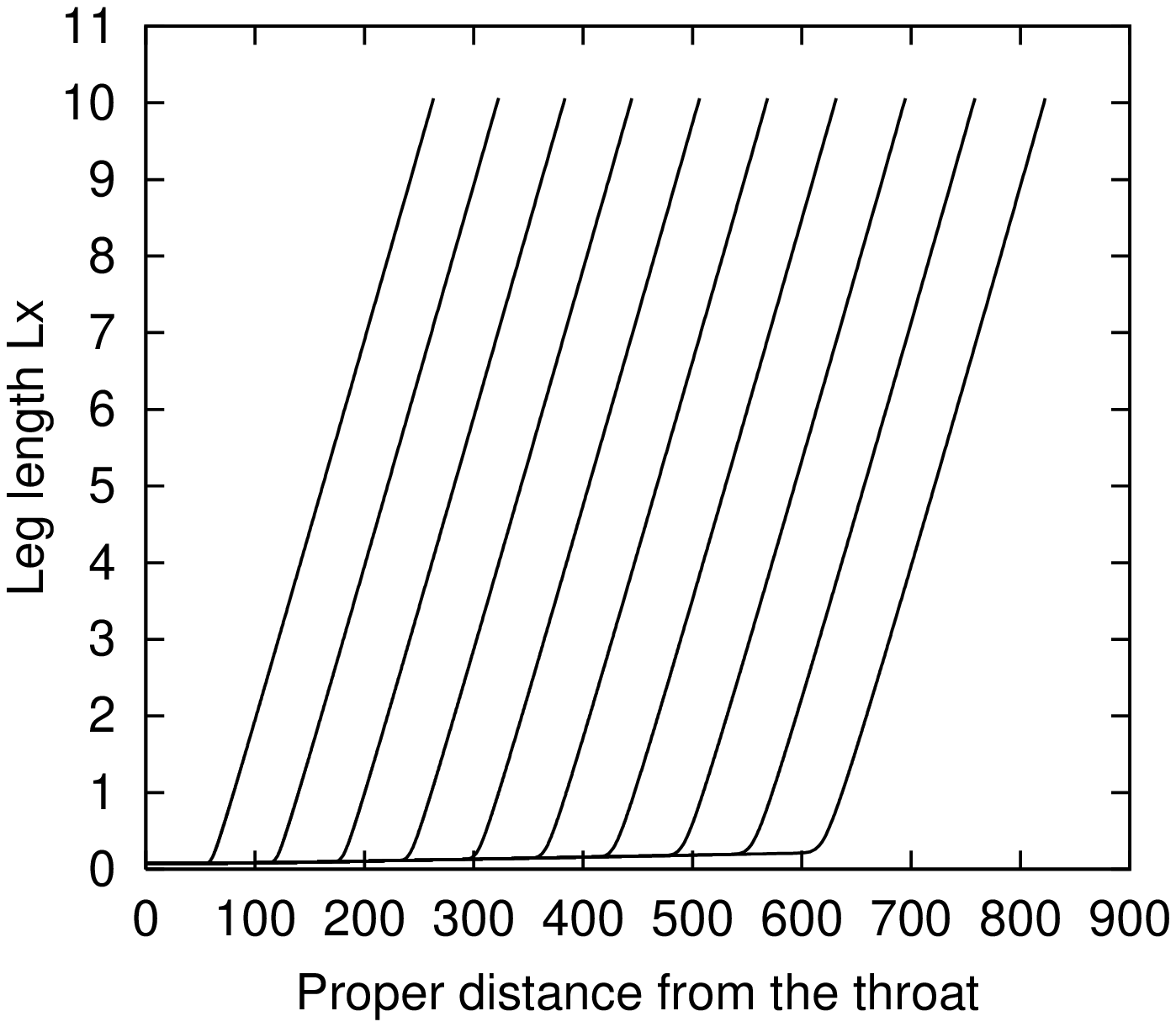,width=5.5cm,height=4.1cm}
    \hskip 1cm
    \epsfig{file=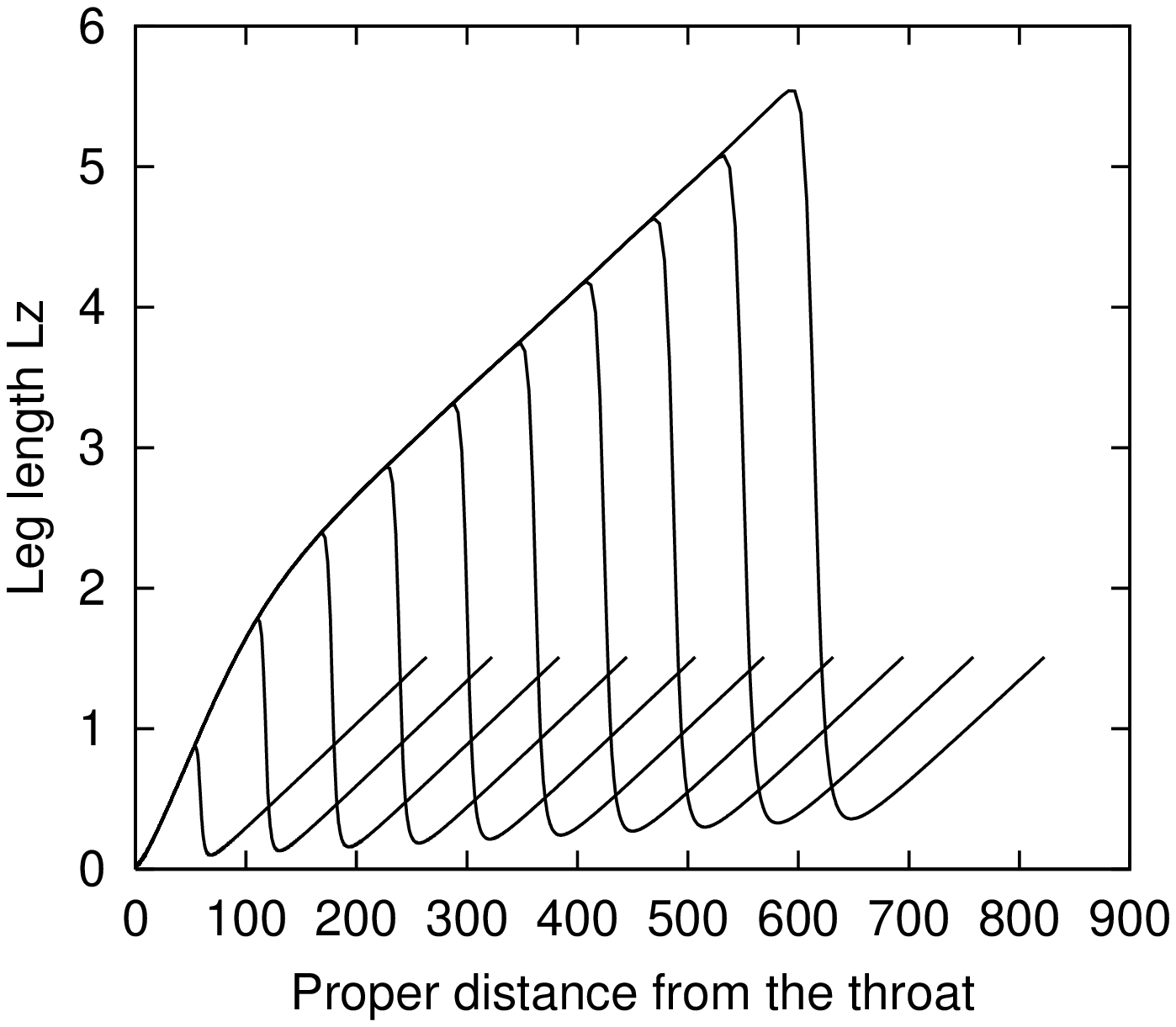,width=5.5cm,height=4.1cm}
    \caption{The leg lengths $L_x$ and $L_z$ versus proper distance from the
    throat at times 100m to 1000m in steps of 100m. Same initial grid as
    Bernstein, Hobill and Smarr but with 800 grid points.}
    \label{fig:figa}
\end{minipage}\\
\begin{minipage}[t]{1.0\textwidth}
\vbox to 0.25cm{\ \vfill}
\end{minipage}
\begin{minipage}[t]{\textwidth}
    \epsfig{file=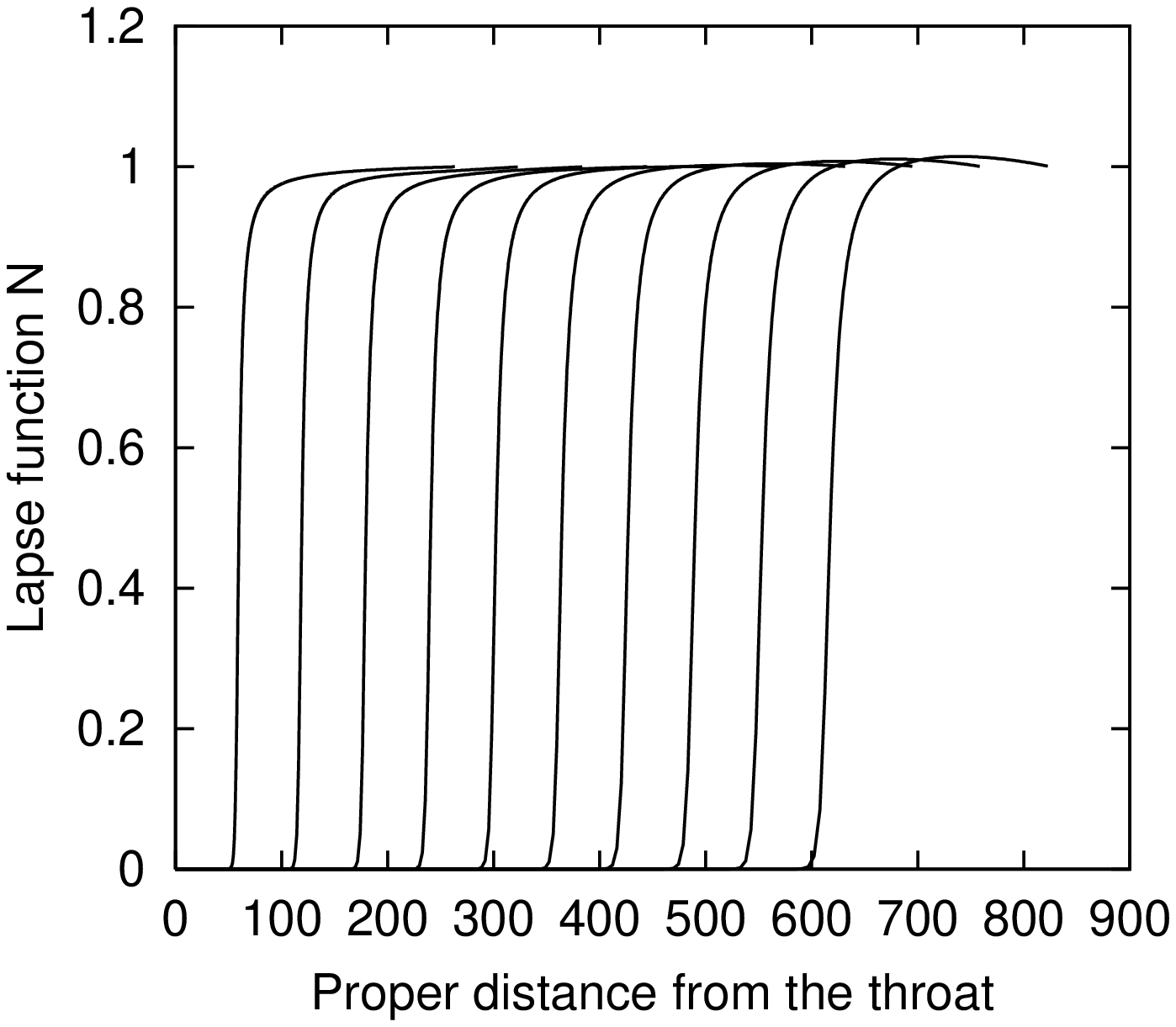,width=5.5cm,height=4.1cm}
    \hskip 1cm
    \epsfig{file=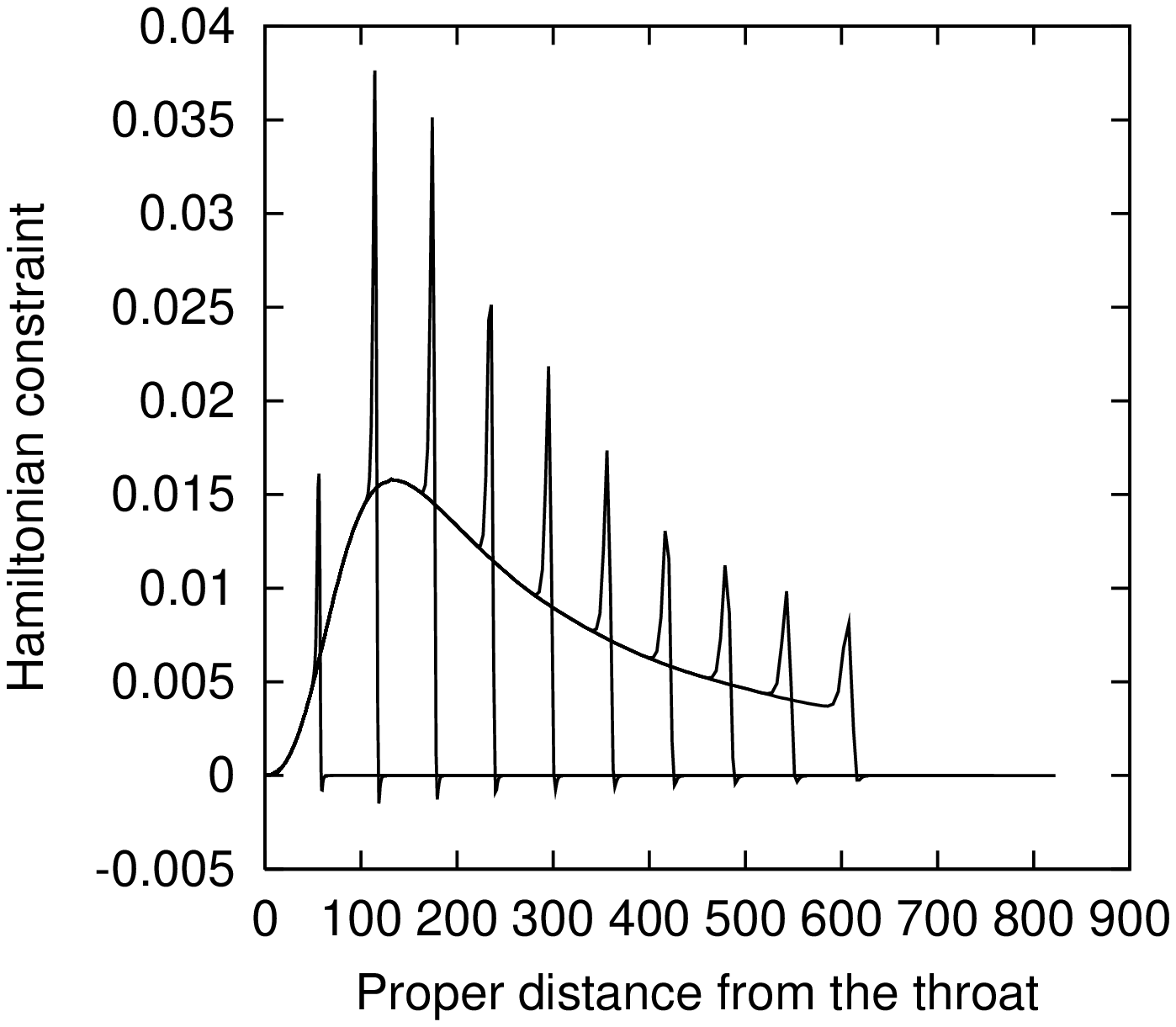,width=5.5cm,height=4.1cm}
    \caption{The lapse and Hamiltonian functions at times 100m to 1000m in steps
    of 100m.}
    \label{fig:figb}
\end{minipage}\\
\end{figure}

Our results are displayed in figures (\ref{fig:figa},\ref{fig:figb}) and they
clearly show no signs of an instability out to $t=1000m$. In contrast, previous
results\cite{Anninos} using maximal slicing would crash at $t\approx100m$.
Though our results are stable there is evidence of loss of accuracy,
particularly in the curvatures, in the later stages of the evolution. We
attribute this to a loss of resolution near the horizon as the lattice becomes
increasingly stretched due to the use of maximal slicing. Work is currently in
progress in applying the smooth lattice method to Brill waves in
2d-axisymmetric spacetimes.

\end{document}